\documentclass[aps,pre,preprint,groupedaddress]{revtex4-1}
\usepackage[T1]{fontenc}
\usepackage{graphicx}
\usepackage{listings}
\usepackage{url}

\begin{document}

\lstset{language=Python}
\lstset{tabsize=4}              
\lstset{frame=lines}
\lstset{basicstyle=\small}
\lstset{showstringspaces=false}   

\title{Python for education: permutations}

\author{A. Kapanowski}

\email[Corresponding author: ]{andrzej.kapanowski@uj.edu.pl}

\affiliation{Institute of Physics, Jagiellonian University, ulica Reymonta 4, 30-059 Cracow, Poland}

\date{\today}

\begin{abstract}
Python implementation of permutations is presented.
Three classes are introduced:
\lstinline|Perm| for permutations,
\lstinline|Group| for permutation groups, and
\lstinline|PermError| to report any errors for both classes.
The class Perm is based on Python dictionaries and utilize cycle notation.
The methods of calculation for the perm order, parity, ranking
and unranking are given.
A random permutation generation is also shown.
The class Group is very simple and it is also based on dictionaries.
It is mainly the presentation of the permutation groups interface
with methods for the group order, subgroups (normalizer, centralizer,
center, stabilizer), orbits, and several tests.
The corresponding Python code is contained in the modules
\lstinline|perms| and \lstinline|groups|.
\end{abstract}


\maketitle

\section{Introduction}
\label{sec:intro}

Python is a programming language that is used by many companies,
universities, and single programmers
\cite{python}.
Some of its key features are: very clear, readable syntax;
high level dynamic data types;
exception-based error handling;
extensive standard libraries and third party modules;
availability for all major operating systems.
Python is sometimes called \emph{executable pseudocode},
because it can be used as a prototyping or 
RAD (rapid application development) language.
On the other hand, it was shown that Python can be used
as the first language in the computer science curriculum
\cite{Zelle}, 
\cite{thinkpython}.

Python can be also used to implement classic algorithms and design
new problem-solving algorithms
\cite{Hetland}. 
Although Python is not as fast as C or Java,
in many cases it may be fast enough to do the job.
It is important how our program \emph{scales} with the input size,
what algorithms are used.
A solid understanding of algorithm design is a crucial problem
and Python stimulates experiments and tests.
Python tools as doctest and unittest can reduce the effort involved 
in code testing
\cite{Arbuckle}, 
\cite{Turnquist}.

In this paper we are interested in computational group theory (CGT)
and permutation (perm) groups algorithms
\cite{Seress_intro}.
Perm groups are the oldest type of representations of groups
and perm groups algorithms are among the best developed parts
of CGT. The methods developed by Sims are at the heart of most
of the algorithms
\cite{Seress}.
Many algorithms have been implemented in GAP,
a system for computational discrete algebra
\cite{gap}. GAP code is written in a high-level, Pascal-like language,
and is freely available.

There is also a Python library for symbolic mathematics called SymPy
\cite{sympy}.
SymPy has Combinatorics Module with perms and perm groups.
SymPy code contains many advanced Python features and may be difficult
to read for a novice programmer. The Combinatorics Module imports
some objects from other modules what is also a disturbing factor.
We would like to present a similar code but self-contained, readable,
and avoiding the most advanced Python features.
A Python code is a considerable part of the paper because
it plays a role of a pseudocode and it also presents
Python best practices.

The paper is organized as follows.
In Section~\ref{sec:basic_notions} basic notions of groups are defined.
In Sections~\ref{sec:perm_iface} and~\ref{sec:perm_class}
the implementation of perms is presented (\lstinline|perms| module)
and some usage examples are given.
Perms are based on Python dictionaries and utilize cycle notation.
In Sections~\ref{sec:group_iface} and~\ref{sec:group_class}
the implementation of perm groups is presented (\lstinline|groups| module).
This implementation is very simple and suitable only for groups
of sufficiently small order.
It is included to familiarize the reader with the perm groups interface.
It is important that almost the same interface can be used
for the advanced implementation of perm groups that
will be published elsewhere.
Conclusions are contained in Section~\ref{sec:conclusions}.

\section{Basic notions of groups}
\label{sec:basic_notions}

A group $G$ is a set together with an operation $*$ that combines
any two elements from $G$ to form another element from $G$
\cite{wiki_group}.
The operation $*$ must satisfy four requirements:

\begin{enumerate}
\item \emph{Closure}.
For all $a,b$ in $G$, $a*b$ is in $G$.
\item \emph{Associativity}.
For all $a,b,c$ in $G$, $(a*b)*c=a*(b*c)$.
\item \emph{Identity element}.
There exists $e$ in $G$, such that for all $a$ in $G$, $e*a=a*e=a$.
\item \emph{Inverse element}. 
For each $a$ in $G$, there exists $\tilde{a}$ in $G$, such that 
$a*\tilde{a}=\tilde{a}*a=e$.
\end{enumerate}
A group $G$ is called \emph{abelian} if $a*b=b*a$ for all $a,b$ in $G$.
A group $G$ is \emph{finite} if the set G has a finite number of elements
(the group order $|G|$ is finite).
In this paper, all groups are finite.

A subset $H$ of $G$ is a \emph{subgroup} of $G$ 
if $H$ is a group together with the operation $*$ from $G$.
$H$ is a \emph{normal subgroup} in $G$ ($H \triangleleft G$)
if $a*b*\tilde{a}$ is in $H$, for all $a$ in $G$, for all $b$ in $H$.

If $S$ is a subset of $G$ then we denote by $\langle S \rangle$
the subgroup generated by $S$.
The \emph{commutator} of $a,b$ in $G$ is $[a,b]=a*b*\tilde{a}*\tilde{b}$.
For subgroups $H,K$ of $G$, the commutator of $H$ and $K$ is defined
as $[H,K]=\langle [a,b]| a \in H,b \in K \rangle$.
The commutator $[G,G]$ is called \emph{derived subgroup} of $G$
and it is always a normal subgroup of $G$.
A group $G$ is \emph{perfect} if $[G,G]=G$.
On the other hand, if $[G,G]$ is trivial, then $G$ is abelian.

\section{Interface for permutations}
\label{sec:perm_iface}

\begin{table}
\caption[Interface for perms.]{
\label{tab:class_perm}
Interface for perms; $p$ and $q$ are perms, $c$ and $d$ are cycles
given as Python tuples or lists.
}
\begin{center}
\begin{tabular}{ll}
\hline\hline
Method name & Short description
\\ \hline
\lstinline|Perm()| &  returns the identity perm    \\
\lstinline|Perm()(*c)(*d)| &  returns a perm from cycles   \\
\lstinline|Perm(data=[0,2,1,3])| &  returns a perm from a list   \\
\lstinline|p.is_identity()| & returns True for the identity perm   \\
\lstinline|~p| & returns the inverse of $p$    \\
\lstinline|p * q| & returns the product $p*q$ as a perm   \\
\lstinline|p == q| & returns True for the same perms    \\
\lstinline|p[k]| & returns the item on the position $k$ in $p$    \\
\lstinline|pow(p,m)|, \lstinline|p**m| & returns the m-th power of $p$  \\
\lstinline|p.support()| & returns a list of integers moved by $p$    \\
\lstinline|p.max()| & returns \lstinline|max(p.support())|    \\
\lstinline|p.min()| & returns \lstinline|min(p.support())|    \\
\lstinline|p.list(size)| & returns $p$ in array form   \\
\lstinline|p.label(size)| & returns the perm string label    \\
\lstinline|p.cycles()| & returns a list of perm cycles    \\
\lstinline|p.order()| & returns the perm order   \\
\lstinline|p.parity()| & returns the parity of $p$ (0 or 1)   \\
\lstinline|p.is_even()| & returns True if $p$ is even    \\
\lstinline|p.is_odd()| & returns True if $p$ is odd    \\
\lstinline|p.sign()| & returns the perm sign ($+1$ or $-1$)   \\
\lstinline|p.commutes_with(q)| & returns True if \lstinline|p*q==q*p|   \\
\lstinline|p.commutator(q)| & returns the commutator $[p,q]$    \\
\lstinline|Perm.random(size)| & return a random perm    \\
\lstinline|p.inversion_vector(size)| & returns the inversion vector of $p$   \\
\lstinline|p.rank_lex(size)| & returns the lexicographic rank of $p$  \\
\lstinline|Perm.unrank_lex(size,rank)| & returns a perm (lexicographic unranking)    \\
\lstinline|p.rank_mr(size)| & returns the Myrvold and Ruskey rank of $p$ \\
\lstinline|Perm.unrank_mr(size,rank)| & returns a perm (M. and R. unranking)   \\
\hline\hline
\end{tabular}
\end{center}
\end{table}

A permutation (perm) is a one-to-one mapping of a set onto itself.
If $p$ and $q$ are perms such that \lstinline|p[i]==j| 
and \lstinline|q[j]==k|, the product \lstinline|(q*p)[i]==k|.
Note that many authors have the opposite convention \cite{Knuth}.
The set of all permutations of any given set $X$ of $n$ elements
forms the symmetric group $\mbox{Sym}(X)$ or $S_n$.
The order of $S_n$ is $n!$.
Any subgroup of a symmetric group $S_n$ is called 
a perm group of degree $n$.

Perms are often shown as an array with two rows
\begin{lstlisting}
[  0    1    2  ...   n-1 ]
[p[0] p[1] p[2] ... p[n-1]]
\end{lstlisting}
Sometimes, only the second line is used 
to present a perm (array form).
Note that we have $X=\{0,1,2,\ldots,n-1\}$.

The third method of notation is cycle notation.
A cycle (k-cycle) $c$ with the length $k$ can be written as a Python 
tuple \lstinline|(c[0],c[1],...,c[k-1])| but, in fact,
a Python list can be also used in programming perms.
It corresponds to the permutation $q$,
where \lstinline|q[c[i]]==c[i+1]| for $0 \le i < k-1$,
\lstinline|q[c[k-1]]==c[0]|. 
If $j$ is not in the cycle then \lstinline|q[j]==j|.
A 2-cycle is called a transposition.
Any permutation can be expressed as a product of disjoint cycles
(1-cycles are often omitted).
Any cycle can be expressed as a product of transpositions
\lstinline|(c[0],c[k-1])(c[0],c[k-2])...(c[0],c[1])|.

Let us show some properties of perms that are listed 
in Table \ref{tab:class_perm}.
Perms are almost always entered and displayed in disjoint cycle notation.
The perm size $n$ is undefined because keys not defined explicitly are equal 
to their values ($p[i]==i$).

\begin{lstlisting}
>>> from perms import *
>>> p,q,r = Perm()(0,1),Perm()(1,2),Perm()(2,3)
>>> p.is_odd()
True
>>> p*q
Perm()(0,1,2)
>>> q*p
Perm()(0,2,1)
>>> p.commutes_with(q)
False
>>> p.commutator(q)
Perm()(0,2,1)
>>> (p*p).is_identity()
True
>>> (p*r).cycles()
[[0,1],[2,3]]
>>> q.list(size=4)
[0,2,1,3]
>>> p[0],p[4],p[8]
(1,4,8)
>>> pow(p*q*r,1234567890)
Perm()(0, 2)(1, 3)
>>> Perm.random(size=5)
Perm()(0,2)(1,4,3)
>>> q.rank_lex(size=4)
2
>>> q.rank_lex(size=5)
6
>>> Perm.unrank_lex(size=4,rank=20)
Perm()(0,3,2)
>>> Perm.unrank_lex(size=5,rank=20)
Perm()(1,4,3)
>>> q.rank_mr(size=5)
99
>>> Perm.unrank_mr(5,20)
Perm()(0,2,1,3,4)
\end{lstlisting}

\section{Class for permutations}
\label{sec:perm_class}

Now we would like to present Python implementation of perms.
The code was tested under Python 2.6.
Let us define the exception \lstinline|PermError| that will
be used to report all problems.

\begin{lstlisting}
class PermError(Exception):
	"""Error in permutations."""
	pass
\end{lstlisting}

A perm is internally a dictionary where missing keys
(\lstinline|p[k] == k|) are created when they are required.
Initially, only the keys with \lstinline|p[k] != k| have to be created.
The code of the Perm class is fairly self-explanatory.
It is inspired by the \lstinline|Cycle| class from SymPy
\cite{sympy} but has the enhanced functionality.
Note that the binary exponentiation algorithm is used
for finding powers of perms. All integer powers are allowed.
String labels will be used in perm groups.

\begin{lstlisting}
class Perm(dict):
	"""The class defining a permutation."""
	def __init__(self, data=None):
		"""Loads up a Perm instance."""
		if data:
			for key,value in enumerate(data):
				self[key] = value
	def __missing__(self, key):
		"""Enters the key into the dict and returns the key."""
		self[key] = key
		return key
	def __call__(self, *args):
		"""Returns the product of the perm and the cycle."""
		tmp = {}
		n = len(args)
		for i in range(n):
			tmp[args[i]] = self[args[(i+1)%n]]
		self.update(tmp)
		return self
	def is_identity(self):
		"""Test if the perm is the identity perm."""
		return all(self[key] == key for key in self)
	def __invert__(self):
		"""Finds the inverse of the perm."""
		perm = Perm()
		for key in self:
			perm[self[key]] = key
		return perm
	def __mul__(self, other):
		"""Returns the product of the perms."""
		perm = Perm()
		# Let us collect all keys.
		# First keys from other, because self can grow up.
		for key in other:
			perm[key] = self[other[key]]
		for key in self:
			perm[key] = self[other[key]]
		return perm
	def __eq__(self, other):
		"""Test if the perms are equal."""
		return (self*~other).is_identity()
	def __getitem__(self, key):
		"""Finds the item on the given position."""
		return dict.__getitem__(self, key)
	def __pow__(self, n):
		"""Finds powers of the perm."""
		if n == 0:
			return Perm()
		if n < 0:
			return pow(~self, -n)
		perm = self
		if n == 1:
			return self
		elif n == 2:
			return self*self
		else:         # binary exponentiation
			tmp = Perm()  # identity
			while True:
				if n % 2 == 1:
					tmp = tmp*perm
					n = n-1
					if n == 0:
						break
				if n % 2 == 0:
					perm = perm*perm
					n = n/2
		return tmp
	def support(self):
		"""Returns the elements moved by the perm."""
		return [key for key in self if self[key] != key]
	def max(self):
		"""Return the highest element moved by the perm."""
		if self.is_identity():
			return 0
		else:
			return max(key for key in self if self[key] != key)
	def min(self):
		"""Return the lowest element moved by the perm."""
		if self.is_identity():
			return 0
		else:
			return min(key for key in self if self[key] != key)
	def list(self, size=None):
		"""Returns the perm in array form."""
		if size is None:
			size = self.max()+1
		return [self[key] for key in range(size)]
	def label(self, size=None):
		"""Returns the string label for the perm."""
		if size is None:
			size = self.max()+1
		if size > 62:
			raise PermError("size is too large for labels")
		letters = "0123456789ABCDEFGHIJKLMNOPQRSTUVWXYZ"
		letters = letters + "abcdefghijklmnopqrstuvwxyz_"
		tmp = []
		for key in range(size):
			tmp.append(letters[self[key]])
		return "".join(tmp)
\end{lstlisting}
Most basic operations require $O(n)$ time for perms from $S_n$.
The binary exponentiation takes $O(n \log(m))$ time for the power $m$.

\subsection{Cycles}

The method \lstinline|cycles()| returns a list of cycles
without 1-cycles. It is used to get the string representation
of a perm and to compute the order of a perm via the functions
\lstinline|lcm()| and \lstinline|gdc()|
\cite{sympy}.
When a perm is raised to the power of its order it equals the identity perm,
\lstinline|pow(p,p.order())==Perm()|.
Note that the code of the method \lstinline|order()|
is exceptionally compact and transparent.

\begin{lstlisting}
def gcd(a, b): 
	"""Computes the greatest common divisor."""
	while b:
		a, b = b, a%b
	return a

def lcm(a, b):
	"""Computes the least common multiple."""
	return a*b/gcd(a, b)

class Perm(dict):
# ... other methods ...
	def cycles(self):
		"""Returns a list of cycles for the perm."""
		size = self.max()+1
		unchecked = [True] * size
		cyclic_form = []
		for i in range(size):
			if unchecked[i]:
				cycle = []
				cycle.append(i)
				unchecked[i] = False
				j = i
				while unchecked[self[j]]:
					j = self[j]
					cycle.append(j)
					unchecked[j] = False
				if len(cycle) > 1:
					cyclic_form.append(cycle)
		return cyclic_form
	def __repr__(self):
		"""Computes the string representation 
		of the perm."""
		tmp = ["Perm()"]
		for cycle in self.cycles():
			tmp.append(str(tuple(cycle)))
		return "".join(tmp)
	def order(self):
		"""Returns the order of the perm."""
		tmp = [len(cycle) for cycle in self.cycles()]
		return reduce(lcm,tmp,1)
\end{lstlisting}

\subsection{Parity}

Every permutation can be expressed as a product of transpositions.
There are many possible expressions for a given perm but 
the parity of the transposition number is preserved.
All permutations are then classified as even or odd,
according to the transposition number.
The set of all even permutations from the symmetric group $\mbox{Sym}(X)$
forms the alternating group $\mbox{Alt}(X)$ or $A_n$.
The order of $A_n$ is $n!/2$.

\begin{lstlisting}
class Perm(dict):
# ... other methods ...
	def parity(self):
		"""Returns the parity of the perm (0 or 1)."""
		size = self.max()+1
		unchecked = [True] * size
		# c counts the number of cycles in the perm including 1-cycles
		c = 0
		for j in range(size):
			if unchecked[j]:
				c = c+1
				unchecked[j] = False
				i = j
				while self[i] != j:
					i = self[i]
					unchecked[i] = False
		return (size - c) % 2
	def is_even(self):
		"""Test if the perm is even."""
		return self.parity() == 0
	def is_odd(self):
		"""Test if the perm is odd."""
		return self.parity() == 1
	def sign(self):
		"""Returns the sign of the perm (+1 or -1)."""
		return (1 if self.parity() == 0 else -1)
\end{lstlisting}

\subsection{Commutators and random perms}

Here we define the commutator of two perms $p,q$ as
\lstinline|p*q*(~p)*(~q)|.
A random perm generator uses the Python \lstinline|random| module.

\begin{lstlisting}
class Perm(dict):
# ... other methods ...
	def commutes_with(self, other):
		"""Test if the perms commute."""
		return self*other == other*self
	def commutator(self, other):
		"""Finds the commutator of the perms."""
		return self*other*~self*~other
	@classmethod
	def random(self, size):
		"""Returns a random perm of the given size."""
		import random
		tmp = range(size)
		random.shuffle(tmp)
		return Perm(data=tmp)
\end{lstlisting}

\subsection{Ranking and unranking permutations}

A ranking function for perms on $n$ elements assigns a unique
integer in the range from 0 to $n!-1$ to each of the $n!$ perms.
The corresponding unranking function is the inverse
\cite{2001_Myrvold_Ruskey}.
The algorithm for ranking perms in lexicographic order uses
the inversion vector and it takes $O(n^2)$ time.
The inversion vector consists of elements whose value indicates 
the number of elements in the perm that are lesser than it and lie 
on its right hand side \cite{sympy}.
The inversion vector is the same as the Lehmer encoding of a perm.

In 2001 Myrvold and Ruskey presented simple ranking and unranking 
algorithms for perms that can be computed using $O(n)$ arithmetic operations
\cite{2001_Myrvold_Ruskey}.
It is inspired by the standard algorithm for generating a random perm.
Myrvold and Ruskey algorithms are shown in functions
\lstinline|rank_mr()| and \lstinline|unrank_mr()|.

\begin{lstlisting}
def swap(L, i, j):
	"""Exchange of two elements on the list."""
	L[i], L[j] = L[j], L[i]

class Perm(dict):
# ... other methods ...
	def inversion_vector(self, size):
		"""Returns the inversion vector of the perm."""
		lehmer = [0]*size
		for i in range(size):
			counter = 0
			for j in range(i+1, size):
				if self[i] > self[j]:
					counter = counter+1
			lehmer[i] = counter
		return lehmer
	def rank_lex(self, size):
		"""Returns the lexicographic rank of the perm."""
		lehmer = self.inversion_vector(size)
		lehmer.reverse()
		k = size-1
		res = lehmer[k]
		while k > 0:   # a modified Horner algorithm
			k = k-1
			res = res*(k+1) + lehmer[k]
		return res
	@classmethod
	def unrank_lex(self, size, rank):
		"""Lexicographic perm unranking."""
		alist = [0]*size
		i = 1
		while i < size:
			i = i+1
			alist[i-1] = rank % i
			rank = rank/i
		if rank > 0:
			raise PermError("size is too small")
		alist.reverse()   # this is the inversion vector
		E = range(size)
		tmp = []
		for item in alist:
			tmp.append(E.pop(item))
		return Perm(data=tmp)
	def rank_mr(self, size):
		"""Myrvold and Ruskey rank of the perm."""
		alist = self.list(size)
		blist = (~self).list(size)   # inverse
		return Perm._mr_helper(size,alist,blist)
	@classmethod
	def _mr_helper(self, size, alist, blist):
		"""A helper function for MR ranking."""
		# both alist and blist are modified
		if size == 1:
			return 0
		s = alist[size-1]
		swap(alist,size-1,blist[size-1])
		swap(blist,s,size-1)
		return s + size*Perm._mr_helper(size-1,alist,blist)
	@classmethod
	def unrank_mr(self, size, rank):
		"""Myrvold and Ruskey perm unranking."""
		tmp = range(size)
		while size > 0:
			swap(tmp, size-1, rank % size)
			rank = rank/size
			size = size-1
		return Perm(data=tmp)
\end{lstlisting}

\section{Interface for permutation groups}
\label{sec:group_iface}

A perm group is a finite group $G$ whose elements are
perms of a given finite set $X$ (usually numbers from 0 to $n-1$)
and whose group operation is the composition of perms
\cite{2007_Joyner_Kohel}.
The number of elements of $X$ is called the degree of $G$.

\begin{table}
\caption[Interface for perm groups.]{
\label{tab:class_group}
Interface for perm groups;
$G$, $H$, and $K$ are groups, $p$ and $q$ are perms.
}
\begin{center}
\begin{tabular}{ll}
\hline\hline
Method name & Short description
\\ \hline
\lstinline|Group()| &  returns a trivial group     \\
\lstinline|G.order()| & returns the group order    \\
\lstinline|G.is_trivial()| & returns True if $G$ is trivial    \\
\lstinline|p in G| & returns True if $p$ belongs to $G$    \\
\lstinline|G.insert(p)| & generates new perms in $G$ from $p$   \\
\lstinline|G.iterperms()| & generates perms from $G$ on demand   \\
\lstinline|G.iterlabels()| & generates perm labels on demand    \\
\lstinline|G.is_abelian()| & returns True if $G$ is abelian    \\
\lstinline|H.is_subgroup(G)| & returns True if $H$ is a subgroup of $G$   \\
\lstinline|H.is_normal(G)| & returns True if $H$ is a normal subgroup of $G$  \\
\lstinline|G.normalizer(H)| & returns the normalizer of $H$ in $G$   \\
\lstinline|G.centralizer(H)| & returns the centralizer of $H$ in $G$  \\
\lstinline|G.center()| & returns the center of $G$   \\
\lstinline|G.orbits(points)| & returns a list of orbits   \\
\lstinline|G.is_transitive(points)| & returns True if $G$ is transitive   \\
\lstinline|G.stabilizer(point)| & returns a stabilizer subgroup   \\
\hline\hline
\end{tabular}
\end{center}
\end{table}

Let us show some computations with perm groups using methods
listed in Table \ref{tab:class_group}.
We will find the relation 
$1 \triangleleft V_4 \triangleleft A_4 \triangleleft S_4$,
where the unity denotes the trivial group 
and $V_4$ is a Klein four-group.

\begin{lstlisting}
>>> from groups import *
>>> s4 = Group()
>>> s4.insert(Perm()(0,1))
>>> s4.insert(Perm()(0,1,2,3))
>>> s4.order()      # the symmetric group S_4
24
>>> a4 = s4.commutator(s4,s4)
>>> a4.order()      # the alternating group A_4
12
>>> all(perm.is_even() for perm in a4.iterperms())
True
>>> a4.is_normal(s4)
True
>>> v4 = s4.commutator(a4,a4)
>>> v4.order()      # the Klein four-group V_4
4
>>> v4.is_abelian()
True
>>> v4.is_normal(a4)
True
\end{lstlisting}

\section{Class for permutation groups}
\label{sec:group_class}

The class \lstinline|Group| is based on Python dictionaries.
All elements of a group are kept,
keys are string labels of perms, values are instances of the
\lstinline|Perm| class.
It is clear that it is possible to handle only small groups
because of the limited computer memory.

\begin{lstlisting}
class Group(dict):
	"""The class defining a perm group."""
	def __init__(self):
		"""Loads up a Group instance."""
		perm = Perm()
		self[perm.label()] = perm
	order = dict.__len__
	def __contains__(self, perm):
		""" Test if the perm belongs to the group."""
		return dict.__contains__(self, perm.label())
	def iterperms(self):
		"""The generator for perms from the group."""
		return self.itervalues()
	def iterlabels(self):
		"""The generator for perm labels from the group."""
		return self.iterkeys()
	def is_trivial(self):
		"""Test if the group is trivial."""
		return self.order() == 1
	def insert(self, perm):
		"""The perm inserted into the group generates new 
		perms in order to satisfy the group properties."""
		label1 = perm.label()
		if perm in self:
			return
		old_order = self.order()
		self[label1] = perm
		tmp1 = {}               # perms added
		tmp1[label1] = perm
		tmp2 = {}               # perms generated
		new_order = self.order()
		while new_order > old_order:
			old_order = new_order
			for label1 in tmp1:
				for label2 in self.iterlabels():
					perm3 = tmp1[label1]*self[label2]
					label3 = perm3.label()
					if perm3 not in self:
						tmp2[label3] = perm3
			self.update(tmp2)
			tmp1 = tmp2
			tmp2 = {}
			new_order = self.order()
	def is_abelian(self):
		"""Test if the group is abelian."""
		for perm1 in self.iterperms():
			for perm2 in self.iterperms():
				if not perm1.commutes_with(perm2):
					return False
		return True
\end{lstlisting}

\subsection{Subgroups}

A group $H$ is a subgroup of a group $G$ if all elements of $H$ belong to $G$.
The \emph{centralizer} of a subset $S$ of $G$ is a set
$C_G(S)=\{ g \in G|s*g=g*s, s \in S \}$
\cite{wiki_centralizer}.
It is clear that $C_G(S)=C_G(\langle S \rangle)$ and that is why
the argument of the method \lstinline|centralizer()| is a group.

The \emph{normalizer} of $S$ in $G$ is a set
$N_G(S)=\{ g \in G|g*s*\tilde{q} \in S, s \in S \}$
\cite{wiki_centralizer}.
We have $N_G(S)=N_G(\langle S \rangle)$ and that is why
the argument of the method \lstinline|normalizer()| is a group.
The centralizer and normalizer of $S$ are both subgroups of $G$.
The centralizer $C_G(S)$ is always a normal subgroup of the normalizer $N_G(S)$.

The \emph{center} of $G$ is a set $Z(G)=C_G(G)$.
The center of $G$ is always a normal subgroup of $G$.
In the case of the abelian group, we get $Z(G)=G$.
On the other hand, sometimes the center can be trivial.

\begin{lstlisting}
class Group(dict):
# ... other methods ...
	def is_subgroup(self, other):
		"""H.is_subgroup(G) - test if H is a subgroup of G."""
		if other.order() % self.order() != 0:
			return False
		return all(perm in other for perm in self.iterperms())
	def is_normal(self, other):
		"""H.is_normal(G) - test if H is a normal subgroup in G."""
		for perm1 in self.iterperms():
			for perm2 in other.iterperms():
				if perm2*perm1*~perm2 not in self:
					return False
		return True
	def subgroup_search(self, prop):
		"""Returns a subgroup of all elements satisfying 
		the property."""
		newgroup = Group()
		for perm in self.iterperms():
			if prop(perm):
				newgroup.insert(perm)
		return newgroup
	def normalizer(self, other):
		"""G.normalizer(H) - returns the normalizer of H."""
		newgroup = Group()
		for perm1 in self.iterperms():
			if all((perm1*perm2*~perm1 in other) 
			for perm2 in other.iterperms()):
				newgroup.insert(perm1)
		return newgroup
	def centralizer(self, other):
		"""G.centralizer(H) - returns the centralizer of H."""
		if other.is_trivial() or self.is_trivial():
			return self
		newgroup = Group()
		for perm1 in self.iterperms():
			if all(perm1*perm2 == perm2*perm1 
			for perm2 in other.iterperms()):
				newgroup.insert(perm1)
		return newgroup
	def center(self):
		"""Returns the center of the group."""
		return self.centralizer(self)
	def commutator(self, group1, group2):
		"""Returns the commutator of the groups."""
		newgroup = Group()
		for perm1 in group1.iterperms():
			for perm2 in group2.iterperms():
				newgroup.insert(perm1.commutator(perm2))
		return newgroup
\end{lstlisting}

The \lstinline|subgroup_search()| method uses a property 
\lstinline|prop| that has to be callable on group elements
and it has to return \lstinline|True| or \lstinline|False|.

\begin{lstlisting}
# Get A_4 from S_4.
>>> a4 = s4.subgroup_search(lambda perm: perm.is_even())
\end{lstlisting}

\subsection{Group action}

If $G$ is a group and $X$ is a set, then a \emph{group action}
of $G$ on $X$ is a function $F:G \times X \rightarrow X$
that satisfies the following two axioms
\cite{wiki_group_action}:

\begin{enumerate}
\item \emph{Identity}
$F(e,x)=x$ for all $x$ in $X$, where $e$ denotes the identity 
element of~$G$.
\item \emph{Associativity}.
$F(g*h,x)=F(g,F(h,x))$ for all $g,h$ in $G$ and all $x$ in $X$.
\end{enumerate}
The \emph{orbit} of a point $x$ in $X$ is the set
$F(G,x)=\{ F(g,x)|g \in G \}$.
There is an equivalence relation defined by saying $x \sim y$ 
if and only if there exists $g$ in $G$ with $F(g,x)=y$.
Two elements $x$ and $y$ are equivalent if and only if their orbits
are the same, $F(G,x)=F(G,y)$.
The group action is transitive if it has one orbit, $F(G,x)=X$.

For every $x$ in $X$, we define the \emph{stabilizer subgroup} of $x$
as a set $\mbox{Stab}_G(x)=\{g \in G|F(g,x)=x \}$.
For finite $G$ and $X$, the orbit-stabilizer theorem states that 
$|F(G,x)|=|G|/|\mbox{Stab}_G(x)|$.

In the case of a perm group $G$ (perms from $S_n$)
we have the standard action $F(p,k)=p[k]$
for $p$ in $G$, $0 \le k \le n-1$.
In our Python implementation, an orbit is a list of points,
where the points ordering is inessential.

Let us analyze the symmetries of a square (the $D_4$ group)
shown in Figure~\ref{square}.
The symmetry group will be constructed from three flips.

\begin{figure}
\begin{center}
\setlength{\unitlength}{20pt}
\begin{picture}(3,3)
\put(0,0){\line(0,1){3}}
\put(1,0){\line(0,1){3}}
\put(2,0){\line(0,1){3}}
\put(3,0){\line(0,1){3}}
\put(0,0){\line(1,0){3}}
\put(0,1){\line(1,0){3}}
\put(0,2){\line(1,0){3}}
\put(0,3){\line(1,0){3}}
\put(0.3,2.3){0}
\put(1.3,2.3){1}
\put(2.3,2.3){2}
\put(0.3,1.3){3}
\put(1.3,1.3){4}
\put(2.3,1.3){5}
\put(0.3,0.3){6}
\put(1.3,0.3){7}
\put(2.3,0.3){8}
\end{picture}
\end{center}
\caption{
\label{square}
A symmetry group for a square is $D_4$.
The elements of the group can be written as perms of integers
from 0 to 8.}
\end{figure}

\begin{lstlisting}
>>> g8 = Group()
>>> g8.insert(Perm()(0,2)(3,5)(6,8))    # horizontal flip
>>> g8.insert(Perm()(0,6)(1,7)(2,8))    # vertical flip
>>> g8.insert(Perm()(1,3)(2,6)(5,7))    # diagonal flip
>>> g8.order()
8
>>> g8.orbits(range(9))
[[0,6,8,2],[1,7,3,5],[4]]
>>> z2 = g8.center()
>>> [perm for perm in z2.iterperms()]
[Perm(),Perm()(0,8)(1,7)(2,6)(3,5)]
\end{lstlisting}

The first orbit contains the points at the corners, the second
points at the edges, and the third contains the center.
The group cannot move a point at a corner onto a point at an edge
or at the center.
The center of the group consists of a half-turn and the identity.

\begin{lstlisting}
class Group(dict):
# ... other methods ...
	def orbits(self, points):
		"""Returns a list of orbits."""
		used = {}
		orblist = []
		for pt1 in points:
			if pt1 in used:
				continue
			orb = [pt1]     # we start a new orbit
			used[pt1] = True
			for perm in self.iterperms():
				pt2 = perm[pt1]
				if pt2 not in used:
					orb.append(pt2)
					used[pt2] = True
			orblist.append(orb)
		return orblist
	def is_transitive(self, points, strict=True):
		"""Test if the group is transitive (has a single orbit). 
		If strict is False, the group is transitive if it has 
		a single orbit of the length different from 1."""
		if strict:
			return len(self.orbits(points)) == 1
		else:         # we ignore static points
			tmp = sum(1 for orb in self.orbits(points) 
			if len(orb)>1)
			return tmp == 1
	def stabilizer(self, point):
		"""Returns a stabilizer subgroup."""
		newgroup = Group()
		for perm in self.iterperms():
			if perm[point] == point:
				newgroup.insert(perm)
		return newgroup
\end{lstlisting}

\section{Conclusions}
\label{sec:conclusions}

In this paper, we presented Python implementation of perms,
the \lstinline|Perm| class (the \lstinline|perms| module), 
based on Python dictionaries. 
It is inspired by the \lstinline|Cycle| 
class from SymPy\cite{sympy} but has the enhanced functionality.
The methods of calculation for the perm order, parity, random perms,
ranking and unranking perms are given.
It is interesting that classic algorithms, such as the Euclidean algorithm
and the binary exponentiation, have found the natural application.

The interface for perm groups is also shown by means of the
\lstinline|Group| class  (the \lstinline|groups| module) 
but the implementation is too simple (and slow) to handle large groups. 
The Python code (\emph{executable pseudocode}) can serve as an introduction
to the group theory and Python programming.
We note that almost the same interface can be used for the advanced
implementation of perm groups based on Sims tables
(to be published elsewhere).





\end{document}